\documentclass[runningheads]{llncs}

\usepackage{graphicx}
\usepackage{amsmath,amssymb}
\usepackage{comment}
\usepackage{alltt}

\usepackage{hyperref}
\usepackage{xcolor}
\definecolor{dark-red}{rgb}{0.4,0.15,0.15}
\definecolor{dark-blue}{rgb}{0.15,0.15,0.8}
\definecolor{medium-blue}{rgb}{0,0,0.5}
\hypersetup{
    colorlinks, linkcolor={dark-red},
    citecolor={dark-blue}, urlcolor={medium-blue}
}

\definecolor{seeblau_20}{RGB}{204, 238, 249}
\definecolor{seeblau_35}{RGB}{166, 225, 244}
\definecolor{seeblau_65}{RGB}{89, 199, 235}  % preferred color 
\definecolor{seeblau_100}{RGB}{0, 169, 224}
\definecolor{seeblau_dark}{RGB}{0, 142, 206}

\definecolor{peach_20}{RGB}{254, 226, 221}
\definecolor{peach_35}{RGB}{254, 207, 199}
\definecolor{peach_65}{RGB}{255, 184, 172}
\definecolor{peach_100}{RGB}{254, 160, 144}  % preferred color 
\definecolor{pach_dark}{RGB}{255, 142, 123}

\definecolor{grau_20}{RGB}{225 226 229}
\definecolor{grau_35}{RGB}{184 188 193}
\definecolor{grau_65}{RGB}{154 160 167}  % preferred color 
\definecolor{grau_100}{RGB}{115 120 126}
\definecolor{grau_dark}{RGB}{77 80 84}

\definecolor{petrol_20}{RGB}{156, 198, 207}
\definecolor{petrol_35}{RGB}{106, 170, 183}
\definecolor{petrol_65}{RGB}{57, 141, 159}
\definecolor{petrol_100}{RGB}{7, 113, 135}  % preferred color 
\definecolor{petrol_dark}{RGB}{3, 95, 114}

\definecolor{seegruen_20}{RGB}{113, 209, 204}
\definecolor{seegruen_35}{RGB}{84, 191, 183}
\definecolor{seegruen_65}{RGB}{10, 163, 152}
\definecolor{seegruen_100}{RGB}{10, 144, 134}  % preferred color 
\definecolor{seegruen_dark}{RGB}{6, 126, 121}

\definecolor{karpfenblau_20}{RGB}{180, 188, 214}
\definecolor{karpfenblau_35}{RGB}{130, 144, 187}
\definecolor{karpfenblau_65}{RGB}{88, 107, 164}
\definecolor{karpfenblau_100}{RGB}{62, 84, 150}  % preferred color 
\definecolor{karpfenblau_dark}{RGB}{50, 67, 118}

\definecolor{pinky_20}{RGB}{243, 191, 203}
\definecolor{pinky_35}{RGB}{236, 160, 178}
\definecolor{pinky_65}{RGB}{230, 128, 152}
\definecolor{pinky_100}{RGB}{224, 96, 126}  % preferred color 
\definecolor{pinky_dark}{RGB}{202, 74, 104}

\definecolor{bordeaux_20}{RGB}{210, 166, 180}
\definecolor{bordeaux_35}{RGB}{188, 122, 143}
\definecolor{bordeaux_65}{RGB}{165, 77, 105}
\definecolor{bordeaux_100}{RGB}{142, 32, 67}  % preferred color 
\definecolor{bordeaux_dark}{RGB}{119, 20, 52}

\definecolor{qtill}{RGB}{255,0,0}
\definecolor{qheiko}{RGB}{0,0,255}

\graphicspath{{./figs/}}
\DeclareGraphicsExtensions{.pdf}

\usepackage{algorithm}
\usepackage{algpseudocode}

\usepackage{pgfplots}
\pgfplotsset{compat=1.11,
    /pgfplots/ybar legend/.style={
    /pgfplots/legend image code/.code={%
       \draw[##1,/tikz/.cd,yshift=-0.25em]
        (0cm,0cm) rectangle (3pt,0.8em);},
   },
}
\DeclareUnicodeCharacter{2212}{−}
\usepgfplotslibrary{groupplots,dateplot}
\usepackage{tikz}
\usetikzlibrary{shapes.geometric}
\usetikzlibrary{arrows.meta,arrows}
\usetikzlibrary{patterns,shapes.arrows}
\usetikzlibrary{tikzmark}
\usetikzmarklibrary{highlighting}
\usepackage{eso-pic}% for scaling the nice plots ;)
\usepackage{environ}

\usepackage{subcaption}
\usepackage[skip=2pt]{caption} % globally reduce space between figure and caption

\tikzset{
  every highlight path/.style={fill=yellow!40},        % default
  highlight red/.style={every highlight path/.style={fill=red!30}},
  highlight blue/.style={every highlight path/.style={fill=blue!20}},
  highlight green/.style={every highlight path/.style={fill=green!30}},
}

\begin{document}
\title{Bimodal Synchronization Performance: \\Why Noise and Sparse Connectivity \\Can Improve Collective Timing}
%\title{Low Connectivity and Individual Mistakes Improve Synchronization in Self-Organized Systems}
%\title{Bimodal Performance in Time Synchronization in Distributed Agents}

\author{Till Aust$^{1}$\orcidID{0000-0003-2863-1341} \and
Tianfu Zhang$^{1,2}$\orcidID{0009-0003-2583-1337} \and
Andreagiovanni Reina$^{1,2,3}$\orcidID{0000-0003-4745-992X} \and Heiko Hamann$^{1,2}$\orcidID{0000-0002-2458-8289}}
\authorrunning{T. Aust et al.}
\titlerunning{Bimodal Synchronization Performance}

\institute{$^{1}$Department of Computer and Information Science, University of Konstanz, Konstanz, Germany\\
$^{2}$Centre for the Advanced Study of Collective Behaviour, University of Konstanz, Konstanz, Germany\\
$^{3}$Department of Collective Behaviour, Max Planck Institute of Animal Behavior, Konstanz, Germany \\
\email{\{till.aust, tianfu.zhang, andreagiovanni.reina, heiko.hamann\}@uni-konstanz.de}}

\maketitle              % typeset the header of the contribution

\begin{abstract}
  %The abstract should briefly summarize the contents of the paper in 150--250 words.
  \noindent%
Pulse-coupled oscillator models inspired by firefly synchronization are widely used to study decentralized time coordination in distributed systems. 
We analyze a discrete-time, discrete-phase firefly-inspired synchronization model and show that collective synchrony emer\-ges only near a critical balance between the quorum threshold (fraction of pulsing neighbors required to trigger a phase update) and the pulse duration (how long agents remain detectable to others).
Within this parameter region, the system exhibits bimodal performance: it either reaches near-perfect synchronization or becomes trapped in stable multi-cluster states, where symmetrically phase-offset subgroups mutually reinforce one another and prevent global synchrony.  
Our analysis shows that reducing connectivity or introducing noise suppresses these low-performance states by breaking such symmetric interactions, indicating that highly connected or noiseless systems are not necessarily optimal for collective synchronization. 
\keywords{pulse-coupled oscillators \and synchronization \and collective dynamics \and two-phase performance \and less is more}
\end{abstract}

\section{Introduction}
Many distributed and decentralized systems, both natural and artificial, must coordinate their behavior without access to a shared global clock. 
In these systems, synchronization cannot be imposed centrally, but must instead emerge through self-organization based on local interactions among agents. 
This problem arises across biological and engineered systems, from the collective flashing of fireflies~\cite{Buck1976} to time coordination in wireless sensor networks~\cite{Ganeriwal2003,Degesys2007} and swarm robotics~\cite{perez2015firefly}. 
A central class of models for studying such processes is pulse-coupled oscillators~\cite{Mirollo1990}, in which agents adjust their internal phase in response to signals received from neighbors. 
Classical results suggest that increasing connectivity generally facilitates synchronization, as agents receive timing information from a larger fraction of the system~\cite{Arenas2008,Strogatz20001}.

Here, we study a discrete-time, discrete-phase firefly-inspired synchronization model. Although the model is abstract and has a deliberately primitive synchronization mechanism, it may offer insights about how to implement lightweight coordination in applications, such as underwater swarm robotics (challenging for radio communication) or ultra-lightweight flying drones (extremely limited resources). The model includes two key mechanisms: a finite flashing duration, which determines how long an agent remains in the signaling state during each cycle, and a quorum threshold, which determines the fraction of flashing neighbors required to trigger a phase update. We show that synchronization performance depends critically on the relationship between these two parameters. Near-perfect synchronization occurs only within a narrow region of parameter space, where the quorum threshold approximately matches the fraction of the cycle spent flashing. 

However, this high-performance region is also where we observe bimodal collective behavior. Within this region, the same parameter settings can yield two qualitatively distinct outcomes across independent runs: near-perfect synchronization or becoming trapped in suboptimal yet stable multi-cluster states.  
We identify the underlying mechanism as symmetry-induced subgroup locking: agents self-organize into multiple synchronized subgroups that are symmetrically spaced in phase space. Through the quorum update rule, these subgroups mutually reinforce their phase offsets, preventing convergence to a single synchronized state and effectively partitioning the system.

Surprisingly, increasing connectivity does not improve performance; instead, highly connected systems can fail to synchronize. 
Our key finding is that removing communication links or introducing noise into agents' clock updates can suppress locked low-performance states and improve collective synchronization. We find that limited or noisy interactions can be beneficial, as previously observed in other swarm robotic scenarios~\cite{Aust:ANTS:2022,Zakir:IROS:2024}. In our model, these interventions improve performance by disrupting the symmetric phase relationships that otherwise sustain multi-cluster locking.

Recent research on multi-robot systems has shown that increasing system size or density can lead to similar bimodal performance distributions, in which the system either operates efficiently or collapses into a degraded state~\cite{Kuckling2024}. While such effects have exclusively been attributed to spatial congestion, our results point to a complementary dynamical mechanism: high interaction density can stabilize symmetric low-performance configurations. This suggests that, across different distributed systems, poor scalability may arise not only from physical obstruction but also from overly regular interaction structures that prevent symmetry breaking and trap the collective dynamics in degraded states.

\section{Related Work}
Synchronization in fireflies has been studied for decades~\cite{Buck1976}. 
Tyrrell et al.~\cite{Tyrrel2006} review this phenomenon and show how its underlying mechanism can be used to design decentralized synchronization protocols for wireless ad hoc networks. 
Building on the Mirollo–Strogatz model~\cite{Mirollo1990}, where oscillators adjust their phase upon receiving pulses from neighbors, the authors derive a synchronization algorithm in which nodes periodically broadcast a common signal and update their clocks upon reception. 
To address practical wireless constraints, they propose a modified transmission strategy with delayed emissions and refractory periods, enabling stable synchronization. 
Simulations show that this bio-inspired protocol can synchronize network nodes within a small number of cycles. 

Lyu~\cite{Lyu2015} introduces a generalized cellular automaton model for discrete-time pulse-coupled oscillators on arbitrary graphs to study synchronization in networks of finite-state agents. 
Each oscillator has a discrete phase that advances at each time step but skips one update when it detects a flashing neighbor behind in phase, allowing lagging oscillators to catch up. 
The authors prove that global synchrony emerges for broad classes of network topologies, including trees and paths, as a function of oscillator period and node degree. 

Greenberg and Hastings~\cite{Greenberg1978} introduced a discrete diffusion model in excitable media to study spatial pattern formation from local interactions. 
Each node cycles through resting, excited, and refractory states; a~resting node becomes excited upon detecting an excited neighbor. 
Kinouchi and Copelli~\cite{Kinouchi2006} studied networks of excitable elements on random graphs and showed that cooperative interactions can generate global propagation of activity, collective oscillations, and synchronized dynamics. 
In contrast, our system uses a discrete quorum-threshold update rule, where agents advance their phase only when the fraction of flashing neighbors exceeds a threshold, rather than relying on continuous pulse responses or pairwise inhibitory interactions. 

Kuckling et al.~\cite{Kuckling2024} provide evidence that multi-robot systems exhibit a bimodal performance regime as a function of robot density. They show that near a critical swarm size, system performance becomes bimodal, with runs either maintaining high performance or collapsing into severe congestion. Hence, systems optimized for peak performance may operate near a critical transition, where small changes in robot density can trigger sudden performance breakdowns.

\section{System Model, Parameters, and Perturbations}

\subsection{Firefly System}

We consider a discrete pulse-coupled oscillator model with quorum-based phase acceleration. Unlike classical firefly-inspired synchronization models based on continuous phase-response functions, agents in our model advance their phase only when the fraction of flashing neighbors exceeds a predefined threshold.

We consider a swarm of $N$ static fireflies (agents) interacting over either a $k$-regular bidirectional random graph, where each agent has exactly $k$~neighbors, or a random geometric graph defined by a communication range~$r$. 
In the geometric case, agents are placed uniformly at random in the unit square, and undirected edges are formed between pairs whose Euclidean distance is less than~$r$.
All agents use the same quorum threshold~$\theta$ and flash duration~$f$.

Each agent $i$ maintains an internal clock~$c_i(t) \in \{0,\dots,C-1\}$ of fixed cycle length~$C$. At each discrete simulation step~$t$, the clock is incremented by one and wraps around modulo $C$. The clock determines the agent's flashing phase.
Flashing is a binary state occupying the final fraction $f$ of the clock cycle:
\[
F_i(t) =
\begin{cases}
1 & \text{if } c_i(t) \bmod C \ge (1-f)C \\
0 & \text{otherwise}
\end{cases}\;.
\]
An agent begins flashing when its clock reaches phase~$(1-f)C$ and remains in the flashing state~$F_i=1$ until the cycle resets. 
When agent $i$ enters the flashing phase, 
%($c_i(t) \bmod C = (1-f)C + 1$), 
it observes the states of its neighbors~$\mathcal{N}_i$. If the fraction of flashing neighbors exceeds the quorum threshold $\theta$, the agent advances its clock by one additional tick. This quorum-based acceleration increases the probability that neighboring agents align their flashing phases over time. 

Agents are initialized with clocks drawn uniformly at random.
Each simulation runs for $T$ time steps. 
We define the maximal amplitude of a run as the largest number of agents flashing simultaneously for any time step $t$: 
\[
A_\text{max}=\max_{t \in T}\frac{1}{N}\sum_{i=1}^{N} F_i(t)\;.
\]
We consider synchronization successful if the maximal amplitude reaches at least 85\%, that is, at least 85\% of agents flash simultaneously at least once during the simulation run. Formally, a run is classified as synchronized if $A_{\text{max}} \geq 0.85$,
or equivalently, if there exists a time $t \leq T$ such that $\sum_{i=1}^{N} F_i(t) \geq 0.85$. 
Runs not meeting this criterion (i.e., $A_\text{max}<0.85$) are classified as asynchronous. 
\begin{figure}[t]
    \centering
    \includegraphics[width=\linewidth, trim=0 0cm 0 0.0, clip]{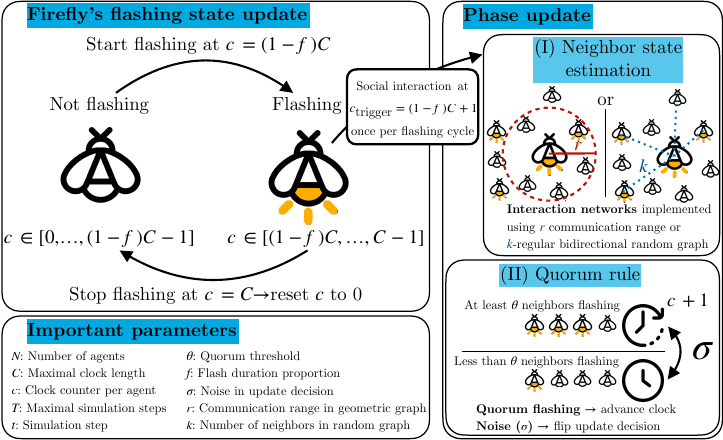}
    \caption{Overview of the firefly-inspired synchronization mechanism. Each agent maintains an internal clock $c$ that is incremented by one at each time step. Once the clock reaches $c=(1-f)C$, the agent enters a flashing state for a duration determined by the flashing fraction~$f$. At the onset of flashing ($(1-f)C+1$), the agent updates its clock according to a local quorum rule: if at least $\theta$ neighboring agents are flashing, its clock is incremented by one additional tick. Interactions are defined either by a range-based communication network with radius~$r$ or by a $k$-regular bidirectional random graph. We introduce noise into the system by either reducing connectivity (reduce number of neighbors) or inverting the update rule with probability~$\sigma$. When~$c=C$, the clock immediately resets to $c=0$, and the cycle repeats.\hfill{\footnotesize
    Credit: Icons are taken from \url{Flaticon.com}.}}
    \label{fig:overview}
\end{figure}

\subsection{Simulation Algorithm}
We implemented the model in Python following the pseudocode described in the electronic supplementary (algorithm~SA1).\footnote{Code and supplementary material: \url{https://github.com/Tianfu-swarm/firefly_two-phase_performance}}
Interaction networks were generated either as random geometric graphs, where neighbors are defined by distance, or as $k$-regular bidirectional random graph using the \textit{igraph} library~\cite{Csardi2006}. 
%Although these network models differ in their degree distributions (Fig.~\ref{fig:avg_neighbors}a), they yield qualitatively similar synchronization behavior (Fig.~\ref{fig:avg_neighbors}b,d). We therefore use $k$-regular graphs for subsequent analyses to isolate the effect of the number of neighbors under fixed homogeneous degree.

%\subsection{System Perturbations}
%We hypothesize that system-level disturbance can be introduced into the swarm either by perturbing the agents' clock update rules or by reducing network connectivity. 
%We want to identify the mechanisms that prevent synchronization. 
%To this end, we hypothesize that introducing system-level noise, either through perturbations to the agents’ clock update rule or through reductions in network connectivity, can break the symmetry and thereby reveal the underlying dynamics governing the lack of synchronization.
To test whether perturbations improve or hinder collective synchronization, we consider two types of disturbances. The first acts on the agents' phase-update decisions by introducing noise into the quorum rule. The second acts on the interaction network by reducing the number of communication links. These disturbances allow us to distinguish whether synchronization failures arise from noisy individual updates or from overly dense and symmetric interaction structures.

\subsubsection{Noise in clock updates. } 
We introduce noise directly into the agents' phase-update decisions. 
Let $M_i(c_{\text{trigger}})$ denote the deterministic quorum decision of agent~$i$ at time step $t$ where the agent's internal clock satisfies~$c_i(t)=c_{\text{trigger}}=(1-f)C+1$: 
\[
M_i(c_{\text{trigger}})=
\begin{cases}
1, & \text{if } \displaystyle \sum_{j\in\mathcal{N}_i} F_j(t) > \theta|\mathcal{N}_i|\\
0, & \text{otherwise}
\end{cases}\;.
\]
If $M_i(c_{\text{trigger}})=1$, the agent advances its clock by one additional step; otherwise, it only performs the regular phase increment. 
To model noisy update decisions, we draw a Bernoulli random variable~$\xi_i(c_{\text{trigger}})\sim \mathrm{Bernoulli}(\sigma)$, 
where the noise level $\sigma\in[0,1]$ is the probability that the deterministic quorum decision is flipped. The noisy update decision is therefore
\[
U_i(c_{\text{trigger}})=
\begin{cases}
M_i(c_{\text{trigger}}) & \text{if } \xi_i(c_{\text{trigger}})=0\\
1-M_i(c_{\text{trigger}}) & \text{if } \xi_i(c_{\text{trigger}})=1
\end{cases}\;.
\]
The update step of agent~$i$ at $c_{\text{trigger}}$ is then
$c_i(t+1)=c_i(t)+1+U_i(c_{\text{trigger}})$,
where the first increment is the regular phase increment at $c_{\text{trigger}}$ and $U_i(c_{\text{trigger}})$ implements the noisy quorum-based acceleration rule (SA.~1, ll.~24-34, shaded red part).

% Noise here refers to doing the opposite action (e.g., when checking for updating and the majority flashes you do increase your internal clock; the majority is not flashing you increase your internal clock) with probability $N_l$. 
% It seems that noise indeed stabilizes the system. This supports our hypothesis that fewer links have the same impact as introducing noise on our decision. 

% Parameters we define \ta{[feel free to add if i missed anything]}: 
% \begin{itemize}
%     \item $C_{\text{flash}}$ is the time interval a firefly is flashing (usually $C_{\text{flash}} = C/2$); $C_{\text{not flash}}$ is the time interval a firefly is not flashing. Further, $C = C_{\text{flash}} + C_{\text{not flash}}$
%     \item $G_i, i \in \mathbb{N}$ subgroups of fireflies that share the same clock
%     \item $H_i, i \in [0, 1, ..., N-1]$ the group of fireflies, firefly $i$ has as neighbors
% \end{itemize}

% Other definitions 
% \begin{itemize}
%     \item We define a successful synchronization as $\exists t \in T: c_{f, t} \geq \frac{C}{2} \, \, \forall f \in N $ or in words: For some time step $t$ all fireflies flash. \ta{Question: do we need to loosen this constraint?}
% \end{itemize}

\subsubsection{Reduced network connectivity. }
We reduce network connectivity by decreasing the degree $k$ of the interaction graph. For each value of $k$, we generate a $k$-regular bidirectional random graph before each experiment, so that every agent has exactly $k$ neighbors, or analogously, in random geometric graphs, we reduce the connectivity range~$r$.

\section{Results}
\label{sec:results}

We first analyze the system behavior across different values of the quorum threshold~$\theta$, proportional flash duration~$f$, and network connectivity.
The parameters~$\theta$ and~$f$ determine, respectively, the fraction of flashing neighbors required for an agent to advance its phase and the fraction of the cycle during which an agent remains in the flashing state. The connectivity determines the number of neighbors available to each agent, which is modulated by either the connectivity range $r$ in the random geometric graphs or the parameter $k$ in the $k$-regular graphs. 
We quantify global synchronization as the maximum amplitude $A_\text{max}$ reached during a simulation of $T=10^3$ time steps with $N=100$ agents and clock-cycle length $C=10$.

\begin{figure}[t]
    \centering
    \includegraphics[width=\linewidth, trim=0 1.1cm 0 0.0, clip]{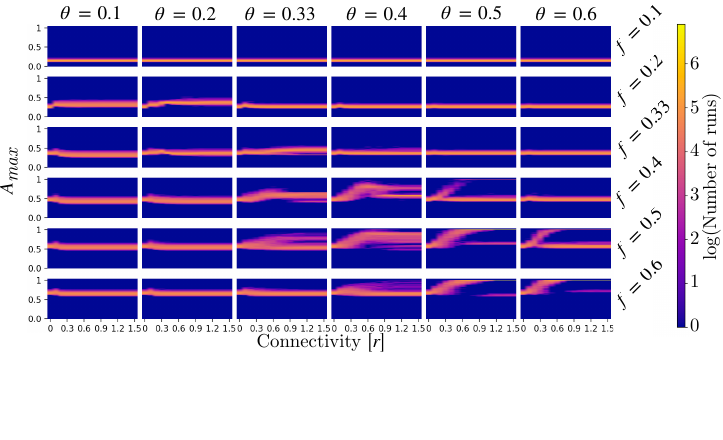}
    \caption{Analysis of quorum threshold~$\theta$ and flash duration~$f$. We find bimodal synchronization effects near the diagonal $\theta \approx f$. 
    The axes of the subplots are the maximal amplitude $A_\text{max}$ (vertical) and connectivity $r$ (horizontal). Higher amplitudes indicate stronger synchronization; colors denote the number of runs attaining each amplitude. A logarithmic color scale is used to make rare outcomes visible.}
    \label{fig:qr_and_fd}
\end{figure}

Figure~\ref{fig:qr_and_fd} shows synchronization performance as a function of connectivity $r$ for different combinations of $\theta$ and~$f$ (a similar analysis for $k$-regular graphs is shown in Supplementary Fig.~SF1, footnote~1). For low connectivity, the interaction network can become sparse or fragmented into disconnected components, making global coordination impossible. Even when the network remains connected, too few interactions can prevent phase information from spreading effectively across the group, and the system generally fails to synchronize. As connectivity increases, synchronization becomes possible, but high connectivity does not lead to uniformly high performance. Instead, the system enters a bimodal regime in which independent runs under the same parameter settings either reach near-perfect synchronization or remain trapped in lower-performance states.

Synchronization performance is highest near $\theta \approx f$, where the quorum threshold approximately matches the fraction of the cycle spent flashing. When the flash duration is short, global synchronization is not achieved, and increasing the quorum threshold alone does not substantially improve performance. 
%When $\theta$ is significantly greater than $f$, an agent triggers a phase advance only when a large proportion of its neighbors are simultaneously in the flashing state. Because the flashing duration is short and the number of agents in the flashing state is limited, it is difficult for most agents to satisfy the quorum condition. The phase difference within the system cannot be effectively reduced and ultimately remains in a highly dispersed state.
%When~$\theta$ is much larger than~$f$, an agent advances only if many neighbors are flashing at the same time. Because flashing is brief, this rarely happens. As a result, phase differences are not reduced effectively. 
%When $f$ is significantly greater than $\theta$, the agent remains in the flashing state for an elongated period. Since the quorum threshold required to trigger an update is also low, different phase clusters may simultaneously satisfy the update conditions. As a result, each subgroup advances in phase simultaneously, while maintaining a phase offset between them. They are unable to gradually merge into a small number of groups with nearly synchronized phases, ultimately resulting in a stable state of subgroup locking.
Conversely, when~$f$ is large relative to $\theta$, agents remain detectable for a long fraction of the cycle while requiring only a small fraction of flashing neighbors to trigger an update. Under these conditions, multiple phase clusters can satisfy the update condition, advance together, and preserve their phase offsets, producing stable multi-group states rather than global synchrony.

\begin{figure}[b]
    \centering
    \includegraphics[width=0.32\linewidth]{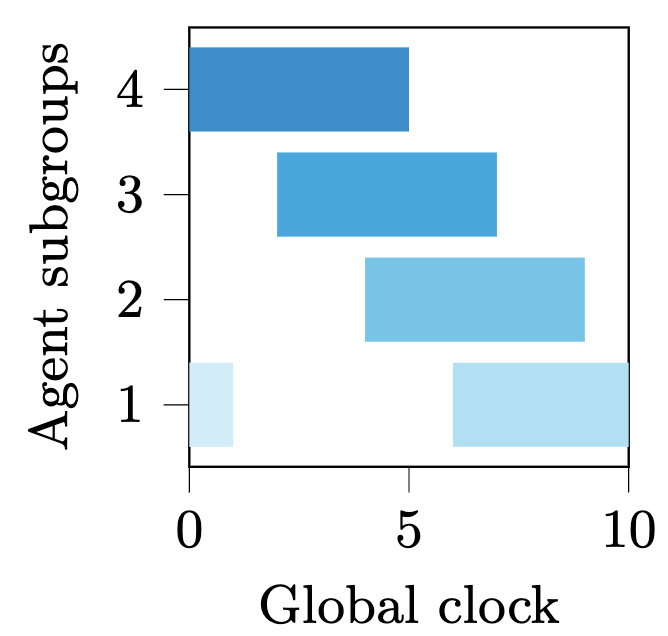}
    \caption{Synchronized subgroups in phase space, where bars indicate the flashing period. A~subgroup is defined as a set of agents with synchronized clocks.}
    \label{fig:failed_scenario}
\end{figure}

These results indicate that synchronization depends on a balance between the availability of flashing signals and the selectivity of the quorum rule. If $\theta$ is much larger than $f$, flashing events are too brief and too sparse to trigger sufficient phase advancement. If $f$ is much larger than $\theta$, the update rule becomes too permissive and can stabilize multiple phase-offset groups. Near $\theta \approx f$, the system enters a bimodal regime: most runs reach near-perfect synchronization, whereas others remain trapped in lower-performance synchronized states. We hypothesize that these low-performance outcomes arise from symmetry-induced subgroup locking and mutual excitation. %(Fig.~\ref{fig:failed_scenario}). 
This symmetric locked-in condition is illustrated in Fig.~\ref{fig:failed_scenario}.
We obtain qualitatively similar results when interactions are defined by $k$-regular graphs rather than random geometric graphs (Supplementary Fig.~SF1, footnote~1). Therefore, in the remainder of the paper, we focus on the balanced regime ($\theta \approx f$),  setting $\theta=0.5$ and~$f=0.5$, and use $k$-regular graphs to isolate the effect of connectivity under fixed homogeneous degree.

% \ta{You can start from here: We report the results for varying quorum-sensing thresholds and flash durations in Fig.~\ref{fig:qr_and_fd}. 
% For small proportional flash durations, the swarm fails to achieve synchronization; similarly, increasing the quorum-sensing threshold alone is insufficient to induce synchronization. 
% In contrast, jointly increasing both parameters enables the swarm to synchronize. 
% However, this improvement comes with a trade-off: the system exhibits bistable or bimodal behavior, where synchronization is achieved in most instances, but certain initial configurations lead to lockstep states that prevent the swarm from synchronization.}

\subsection{Breaking Symmetry through Noise and Topology Changes}
To investigate the role of symmetry-induced subgroup locking, we perturb the system through noisy clock updates and random link removal. 
While noisy clock updates directly disrupt symmetric phase advancement, link removal breaks the symmetry in the number of flashing neighbors perceived by different agents and subgroups. Both types of perturbations can therefore destabilize locked phase configurations and allow the system to escape low-performance states.

\begin{figure}[t]
    \centering
    \includegraphics[width=\linewidth, trim=0 1.1cm 0 0.0, clip]{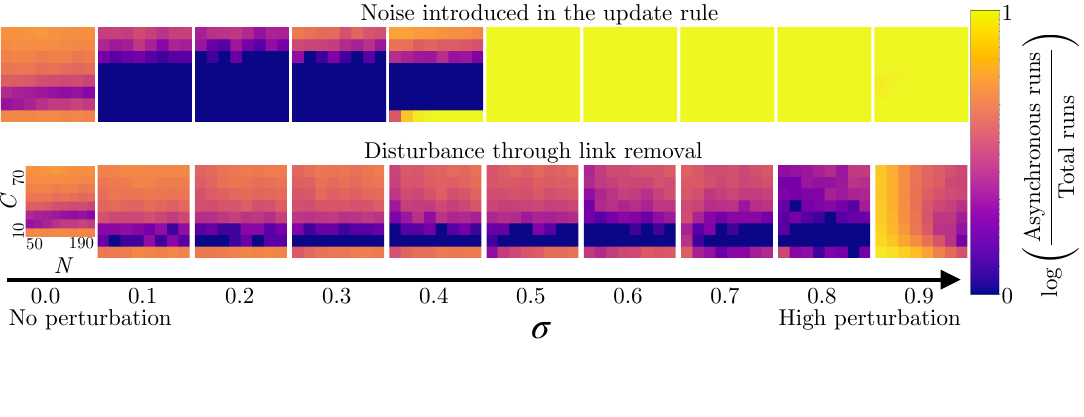}

    \caption{Pushing the system out of low-sync states through increasing perturbations (across panels from left to right). We vary swarm size~$N$ (x-axis) and cycle length~$C$ (y-axis). Blue denotes well-synchronized runs ($A_\text{max}>$85\% simultaneous flashing) and yellow denotes insufficiently synchronized runs ($A_\text{max}\leq$85\%), aggregated over 1000 independent runs ($T=10^4$, $\theta=0.5$, $f=0.5$). The top row introduces noise via the clock update rule, while the bottom row applies disturbance through stochastic link removal while preserving $k$-regularity.}
    \label{fig:introducing_noise}
\end{figure}
\subsubsection{Noisy clock updates. }
Figure~\ref{fig:introducing_noise}~(upper row) shows the effect of clock update noise~$\sigma$ under full connectivity ($k=N-1$) across swarm size~$N \in [50,190]$ and clock cycle length~$C \in [10,70]$. 
Each configuration is evaluated over 1000~independent trials and classified by its maximum synchronization level: blue indicates successful runs, with more than 85\% of agents flashing simultaneously, whereas yellow indicates runs below this threshold. Therefore, a higher prevalence of yellow indicates a low-sync states that persist despite noise. 

Without noise in the clock-advancing mechanism (Fig.~\ref{fig:introducing_noise}, upper left), low-sync runs (warmer colors) dominate for almost all swarm sizes~$N$ and clock cycle lengths~$C$. 
By noise~$\sigma$ increases (Fig.~\ref{fig:introducing_noise}, upper row, left to right), larger regions of well-synchronized runs appear (blue), confirming that noisy updates can break the symmetry of the locked subgroups. 
For $0.1 \leq \sigma \leq 0.4$, low-sync runs remain more frequent at larger clock cycle lengths $C$. This occurs because synchronization becomes more difficult when agents are more dispersed across the clock cycle, leaving a larger number of available phase slots in which persistent subgroups can form instead of collapsing into global synchrony.
For higher noise levels (Fig.~\ref{fig:introducing_noise}, upper row, middle to right), synchronization again breaks down across all tested swarm sizes~$N$ and cycle lengths~$C$. 
In this regime, update decisions become dominated by noise. For $\sigma>0.5$, the deterministic quorum rule is more often inverted than followed, counteracting the synchronization mechanism and driving the system toward active desynchronization.

\subsubsection{Reducing connectivity. }
We then test whether low-sync states can also be destabilized by changing the interaction topology. To isolate this effect, use connected $k$-regular random graphs, so that each agent has exactly $k$ reciprocal neighbors. Connectivity is controlled by the link-removal parameter $\sigma$, which sets the degree to $k = N-1-\lfloor \sigma N \rfloor$ relative to the fully connected case. For each run, we generate a new connected graph and initialize clocks uniformly at random.  
Figure~\ref{fig:introducing_noise}~(lower row) uses the same phase classification as above and shows results across swarm sizes $N$, clock cycle lengths $C$, and link-removal level~$\sigma$.
Each configuration is evaluated over 1000 independent runs of $T=10^4$~steps. 
For $\sigma = 0$, the setup is identical to the noiseless fully connected case shown in the upper row, and therefore produces the same result. 
For increasing link removal, larger and more stable regions of well-synchronized runs emerge up to $\sigma = 0.8$. This indicates that reducing connectivity can break the symmetric interactions that sustain subgroup locking.
However, the effect is weaker and less widespread than that produced by noise in the update rule. 
For $C = 10$, only $\sigma = 0.8$ introduces sufficient perturbation to break subgroup locking. 
For $\sigma = 0.9$, synchronization largely disappears,  indicating that too many links have been removed for the system to support global coordination.

For sparse connectivity in the $k$-regular bidirectional random graph ($k \leq 20$), which occurs, for example, under high link-removal settings ($\sigma = 0.9$), synchronization depends sensitively on the parity of $k$.
For small odd values $k=\{5, 7, 9,\dots, 19\}$, synchronization is difficult and often not achieved (Fig.~\ref{fig:introducing_noise}, bottom right), whereas even values $k=\{6,8,10,\dots,20\}$ lead to the highest global synchrony (Supplementary Fig.~SF2). 
This suggests a possible parity effect in the quorum rule: for $\theta=0.5$, odd and even neighborhood sizes differ in the exact number of flashing neighbors required to trigger an update, which may affect how easily symmetric phase-locked configurations are destabilized. 
%between odd and even $k$: odd values yield a perfectly symmetric majority condition ($P(\text{neighbors flashing} > k/2) = 0.5$), potentially creating metastable coin-flip-like configurations that are difficult to escape, whereas even values weakly break this symmetry. 
Since this interpretation is currently empirical rather than mechanistic, we leave a systematic analysis for future work. We therefore report the more conservative odd-$k$ case in the main text and provide the even-$k$ results, which show stronger synchronization, in the Supplementary Material (Fig.~SF2).

\section{Discussion}
\label{sec:discussion}
We discuss the mechanisms underlying the bimodal synchronization behavior observed in our model and relate them to broader findings on scalable distributed systems. We first explain how symmetric subgroup locking can produce persistent low-performance states, and then discuss how noise or reduced connectivity can destabilize these deadlock states and restore synchronization.

\subsection{Causes of Bimodal Synchronization Quality}

Under all-to-all communication and noiseless decision-making, one might expect synchronization to be guaranteed, since each agent has access to the flashing state of the entire swarm. 
Counterintuitively, however, synchronization can fail. 
This failure depends on the initial phase distribution. For certain distributions (e.g., phases initialized uniformly across the cycle), the swarm does not converge to a single synchronized cluster but instead self-organizes into multiple subgroups with systematically shifted phases (see Fig.~\ref{fig:failed_scenario}). 
These phase shifts are self-reinforcing. When a subgroup evaluates whether to advance its clock, the other phase-shifted subgroups can provide a quorum-qualified fraction of flashing agents~($>\theta N$), hence, satisfying the threshold condition, while at least one subgroup is not. 
As a result, each subgroup advances while preserving the relative offset between groups. The inter-group differences are therefore not resolved, leaving the swarm permanently partitioned into stable synchronized subgroups. 
This mechanism generalizes beyond all-to-all communication: analogous failure modes arise in high-connectivity regimes, although they occur less frequently as connectivity decreases.

Introducing noise into the clock-advancing mechanism, or by removing communication links, increases the probability of converging to a globally synchronized state. 
When update decisions are noisy, or when agents receive incomplete information because of limited connectivity, the perfect phase alignment that sustains the symmetric subgroups is disrupted. 
The systematic phase offsets that would otherwise lock the swarm into a partitioned state are destabilized, and the system's self-organizing properties drive the swarm toward full synchronization. 
Although noise and link removal act on different aspects of the system, both interventions have a similar functional effect: they introduce sufficient asymmetry into the interaction dynamics to break deadlocked low-performance states.

\subsection{Relation to Bimodal Performance in Multi-robot}
% This bimodal performance in synchronization is the first report of such an effect, where the shared resource is not space. 
% Previous studies~\cite{Kuckling2024} largely attribute performance degradation in multi-robot systems to contention for shared physical space, where increasing robot density leads to traffic congestion, blocking, and reduced throughput. Here, however, the shared resource is the communication channel limited by the discretized interval of oscillation phase. 
% In this way, the system of fireflies is different because an infinite number of agents can populate each discrete phase. However, the total number of groups that can form is limited. 
% A~finding for multi-robot systems was a hysteresis effect~\cite{Kuckling2024}, that is, collapsing from a high-performance state to a low-performance state would not immediately be repaired, for example, by removing a robot from a traffic jam. 
% For our firefly pulse-coupled oscillator model here, we do not observe hysteresis. Once the system is synchronized, adding or removing links does not break synchronization. 

The bimodal synchronization behavior observed here is analogous to bimodal performance regimes recently reported in multi-robot systems~\cite{Kuckling2024}. There, increasing robot density can produce two qualitatively different outcomes: maintaining high throughput and operating efficiently, or collapsing into a low-performance state characterized by congestion, blocking, or traffic jams. Thus, beyond a certain density, adding more agents does not merely reduce performance gradually; it increases the probability of a collective failure mode.

Our system exhibits similar high- and low-performance behavior, but the form of competition differs. In multi-robot systems, agents compete for access to physical space. In our synchronization model, agents do not exclude one another spatially; instead, the relevant competition occurs through phase-dependent interactions. Multiple subgroups can occupy distinct regions of the discrete phase cycle and mutually reinforce their offsets through the quorum rule, preventing convergence to a single synchronized cluster.

This analogy suggests that bimodal performance can arise in distributed systems when increased interaction density stabilizes collective configurations that are difficult to escape. However, the underlying mechanism is likely different in the two cases. In spatially congested multi-robot systems, bimodality is associated with hysteresis, whereas we do not observe hysteresis in our model: once global synchronization is reached, changing connectivity does not by itself destroy it. In summary, our results point to a related phenomenology, but one driven by symmetry-induced locking in phase space rather than by persistent physical congestion.

\section{Conclusion}
We studied a discrete firefly-inspired synchronization model and showed that increasing connectivity does not necessarily improve collective performance. 
Instead, the system may exhibit a bimodal performance distribution, in which independent runs under the same parameter settings either converge to near-perfect synchronization or become trapped in stable multi-cluster states.
We identify symmetry-induced subgroup locking as the key mechanism behind these failures: in highly connected, noiseless settings, symmetrically phase-offset clusters can reinforce one another via the quorum rule, preserving their relative phase differences and preventing global synchronization. 
Counterintuitively, full information can therefore hinder coordination by stabilizing low-performance collective states.
We further show that introducing noise into the update rule or reducing connectivity can resolve this failure mode. Both interventions disrupt the symmetry that sustains subgroup locking, allowing the system to escape low-performance states and recover global synchronization. 
These results highlight that communication or decision-making imperfections are not necessarily detrimental: under some conditions, they can improve collective performance by preventing overly symmetric interaction structures from becoming locked.

Our results suggest that bimodal performance may be a general feature of scalable distributed systems. It can arise not only from spatial congestion~\cite{Kuckling2024}, but also from overly regular interaction structures that stabilize collective states from which the system cannot easily escape. 
Future work should investigate whether the observed phenomenon is specific to this specific pulse-coupled oscillator model and synchronization mechanism or generalizes to broader classes of synchronization models and distributed coordination.

\begin{credits}
\subsubsection{\ackname} This work has been supported by the DFG under Germany's Excellence Strategy, EXC 2117-422037984.

\subsubsection{\discintname} The authors declare that they have no conflicts of interest.
\end{credits}

\footnotesize
\bibliographystyle{splncs04}
\bibliography{main}

\end{document}